\documentclass[twocolumn,showpacs,preprintnumbers,amsmath,amssymb]{revtex4}

\begin{document}

\title{\bf ANGULAR MOMENTUM OF PHOTONS \\ EMITTED BY ATOMS}

\author{M.A. Can$^1$, \"{O}. \c{C}ak{\i}r$^1$, A. Horzela$^2$, E.
Kapuscik$^2$, A.A. Klyachko$^1$, and A.S. Shumovsky$^1$}

\address{$^1$Physics Department, Bilkent University, Bilkent,
Ankara, 06800 Turkey \\ $^2$H.  Niewodnicza\'nski Institute of
Nuclear Physics, ul. Radzikowskiego 152, 31 342 Krakow, Poland}

\begin{abstract}
It is shown that the spin and orbital angular momentum of electric
dipole photons have the same operator structure and may differ
from each other only by spatial dependence in the very vicinity of
the atom. It is shown that the photon twins created by a dipole
forbidden transition can manifest the maximum entanglement with
respect to the angular momentum. It is shown that the states of
photons with projection of angular momentum $m=0$ are less stable
than those with $m= \pm 1$.
\end{abstract}


\maketitle

Recently the problem of angular momentum (AM) of photons has
attracted a great deal of interest (e.g., see
\cite{1,2,3,4,5,6,7,8} and references therein), especially in the
context of quantum information processing \cite{3,4,5,6,7,8,9}.
Usually AM of photons is considered as consisting of the {\it
spin} and {\it orbital} angular momentum (OAM) parts \cite{10}
\begin{eqnarray}
\vec{M} & = & \frac{1}{4 \pi c} \int \vec{r}\times ( \vec{E}
\times \vec{B})d^3r= \vec{S}+ \vec{L}, \label{1} \\ \vec{S} & = &
\frac{1}{4 \pi c} \int ( \vec{E} \times \vec{A})d^3r, \label{2} \\
\vec{L} & = & \frac{1}{4 \pi c} \sum_{\alpha =x,y,z} \int
E_{\alpha}(i \vec{{\mathcal{M}}})A_{\alpha}d^3r. \label{3}
\end{eqnarray}
Here $ \vec{{\mathcal{M}}} \equiv -i \vec{r} \times \vec{\nabla}$
is the quantum mechanical OAM operator and $ \vec{E}, \vec{B},
\vec{A}$ denote the electric field, magnetic induction, and vector
potential. Within the quantum domain, these objects are
represented by linear forms in photon operators of creation and
annihilation, so that (1), (2), and (3) are quadratic forms with
respect to photon operators. In this case, the integrands in
(1)-(3), representing the densities of the corresponding
quantities, are the operators defined in the real space
${\mathbb{R}}^3$ and Hilbert-Fock space of photons.

It is known that a photon has spin 1, but can be observed in only
two spin states (helicities) because of the requirement of
Pincar\'{e} invariance on the light cone \cite{11}. This means
that the AM operator (1) is specified by the dynamic symmetry
group $G=E(2) \times SU(2)$. Here $E(2)$, describing the photon
spin symmetry, is Winger's little group, that is the Euclidean
group, which is a semi-direct product of $SO(2)$ and $T(2)$ - the
group of translations in two dimensions. In turn, $SU(2)$ in $G$
describes the symmetry of OAM.

Nevertheless, consideration of specific models can lead to a
different results. For example, a monochromatic plane wave (both
classical and quantum), travelling in the $z$-direction, has no AM
about the $z$-axis because the Poynting vector $\vec{E} \times
\vec{B}$ is parallel to this axis. At the same time, it has spin
(2) directed along the $z$-axis. The rejection of monochromaticity
leads to the following commutation relations \cite{12}
\begin{eqnarray}
[S_{\alpha},S_{\beta}]=0, \quad [L_{\alpha},L_{\beta}]=i
\epsilon_{\alpha \beta \gamma} (L_{\gamma}+S_{\gamma}), \nonumber
\\ \left[L_{\alpha},S_{\beta} \right]=i \epsilon_{\alpha \beta
\gamma}S_{\gamma}, \label{4}
\end{eqnarray}
corresponding to even more degenerated group $G'=T(3) \times
SU(2)$.

The aim of this work is to examine the structure of the total AM
and its spin and OAM parts in the representation of {\it spherical
waves} that corresponds to real photons are emitted by the atomic
and molecular transitions \cite{13,14}. The wave functions of
spherical photons are represented by vector spherical harmonics,
which are the linear combinations of ordinary spherical harmonics
(eigenstates of OAM operator) and spin states \cite{13}. This
assumes the $SU(2) \times SU(2)$ structure of the total AM of a
photon.

We consider the densities of spin and OAM operators in the
representation of spherical (multipole) photons (integrands in (2)
and (3)). We show that these operators have the same structure in
terms of photon operators but different spatial dependence at
short and intermediate distances. In the wave (far) zone, this
difference vanishes, so that there is no way to distinguish
between the spin and OAM parts of the total AM of a multipole
photon. Besides that, we consider generation of photon twins
entangled with respect to the total angular momentum.

As an example of some considerable interest, let us investigate an
electric dipole ($E1$) transition between the excited $|e;m
\rangle$ and ground $|g \rangle $ states with the AM $j_e=1$ and
$j_g=0$, respectively. Since the angular momentum $j_e=1$ has the
three projections $m=0, \pm 1$, the excited atomic state $|e;m
\rangle$ is triple degenerated. Assume first that such a two-level
atom is located at the center of an ideal spherical cavity of high
radius $R$ (the multipole Jaynes-Cummings model \cite{15}). Then,
the transition
\begin{eqnarray}
|e;m \rangle \rightarrow |g \rangle \nonumber
\end{eqnarray}
gives rise to a monochromatic $E1$ photon with projection $m$ of
AM. The operator vector potential, describing this photon, has the
form \cite{13,14}
\begin{eqnarray}
\vec{A}_k(\vec{r})= \sum_m \vec{\mathcal{A}}_{km}(\vec{r})a_{km}
+H.c. \nonumber \\ = \sum_m N(k)[j_2(kr)\vec{Y}_{1,2,m}- \sqrt{2}
j_0(kr)\vec{Y}_{1,0,m}]a_{km}+H.c., \label{5}
\end{eqnarray}
where $\vec{Y}_{j \ell m}$ is the vector spherical harmonics
\cite{13}, $N(k)= \sqrt{4 \pi \hbar c/(3kV)}$ is the normalization
factor, $V=4 \pi R^3/3$ denotes the cavity volume, $j_{\ell}(kr)$
is a function proportional to the spherical Bessel function but
normalized by the condition
\begin{eqnarray}
\int_0^R j_{\ell}(kr)j_{\ell}(k'r)r^2dr=V \delta_{kk'} , \label{6}
\end{eqnarray}
and $a_{km}$ are the photon operators
\begin{eqnarray}
[a_{km},a^+_{k'm'}]= \delta_{kk'} \delta_{mm'} . \label{7}
\end{eqnarray}
Taking into account that $\vec{E}=- \partial \vec{A}/(c \partial
t)$, we can conclude that the integrands in (2) and (3) contain
the operator constructions of the form $a^+_ma_{m'}$. This means
that, in contrast to energy, there is no vacuum oscillations of
AM.

To specify the spin and OAM at any distance $r$ from the atom, we
take into account that the photon localization appears in the
natural way in the form of wavefront \cite{16} (concerning photon
localization also see \cite{9,17,18} and references therein).
Therefore, to find AM carried by a photon at distance $r$ from the
atom, we should perform integration over the spherical shall of
radius $r$, surrounding the source. Finally, we get
\begin{eqnarray}
\vec{S}(r)=f_S(kr)\vec{J}, \quad \vec{L}(r)=f_L(kr)\vec{J}.
\label{8}
\end{eqnarray}
Here $\vec{J}$ denotes the operator of the total AM in the whole
volume of quantization, having the components
\begin{eqnarray}
\left\{ \begin{array}{ll} J_x = & [a_{k0}^+(a_{k+}+a_{k-}) +H.c.]
/
\sqrt{2} \\ J_y= & i[a^+_{k0}(a_{k+}-a_{k-})-H.c.]/ \sqrt{2} \\
J_z= & a^+_{k+}a_{k+}-a^+_{k-}a_{k-} \end{array} \right. \label{9}
\end{eqnarray}
so that
\begin{eqnarray}
[J_{\alpha},J_{\beta}]=i \epsilon_{\alpha \beta \gamma}J_{\gamma}.
\label{10}
\end{eqnarray}
Thus, Eqs. (9) give a representation of the $SU(2)$ subalgebra in
the Weyl-Heisenberg algebra of photon operators (7).

The distance-dependent functions in (8) have the form
\begin{eqnarray}
f_S(kr) & = & \frac{\hbar}{3V} \left[ 2j_0^2(kr)- \frac{1}{2}
j^2_2(kr) \right] , \nonumber \\ f_L(kr) & = & =\frac{\hbar}{3V}
\frac{3}{2} j^2_2(kr). \nonumber
\end{eqnarray}
It follows from the normalization condition (6) that
\begin{eqnarray}
\int_0^R f_S(kr)r^2dr= \int_0^R f_L(kr)r^2dr= \frac{\hbar}{2},
\nonumber
\end{eqnarray}
so that the total AM operator (1) takes the form
\begin{eqnarray}
\vec{M}= \hbar \vec{J} \nonumber
\end{eqnarray}
as all one can expect for $E1$ photon with AM equal to one. The
total spin and OAM operators have the form
\begin{eqnarray}
\vec{S}=\vec{L}= \frac{\hbar}{2} \vec{J}. \nonumber
\end{eqnarray}
The structure (9) of the AM operator can also be obtained in a
different way through the use of conservation of the total AM in
the process of atom-photon interaction \cite{1,9}.

Thus, spin and OAM of $E1$ photon have the same operator
structure. In particular, this means that the commutation
relations for the components of the density operators in (2) and
(3) take the form
\begin{eqnarray}
{[} S_{\alpha}(r),S_{\beta}(r) {]} & = & i
\epsilon_{\alpha \beta \gamma} f_S(kr) S_{\gamma}(r), \nonumber \\
{[} L_{\alpha}(r),L_{\beta}(r) {]} & = & i
\epsilon_{\alpha \beta \gamma} f_L(kr) L_{\gamma}(r), \nonumber \\
{[} L_{\alpha}(r),S_{\beta}(r) {]} & = & i \epsilon_{\alpha \beta
\gamma} f_L(kr)S_{\gamma}(r), \label{11}
\end{eqnarray}
in contrast to (4) and (5).

It is seen that the spin and OAM density operators have different
spatial dependence at short distances from the atom. Since
\begin{eqnarray}
\lim_{x \rightarrow 0} j_{\ell}(x)= \left\{ \begin{array}{ll} 1, &
\mbox{if $\ell =0$} \\ 0, & \mbox{otherwise} \end{array} \right.
\nonumber
\end{eqnarray}
$f_L(kr)$ vanishes at $kr \rightarrow 0$. Thus, at the very
vicinity of the atom, the photon has only spin, while OAM arises
in the process of propagation. A more detailed investigation shows
that spin density strongly prevails over OAM density at $r<0.1
\lambda$, where $\lambda = 2 \pi /k$ is the wavelength. Since the
maximum of $f_S(kr)$ corresponds to $kr=0$, it is possible to say
that atom creates the photon with spin and without OAM. In turn,
OAM achieves maximum at $r \sim \lambda /2$ (intermediate zone).
It is also seen that the main contribution into the total AM comes
from the near zone in contrast to the energy that derives its main
contribution from the wave zone \cite{19}.

Let us stress that the results for monochromatic multipole photons
in the near and intermediate zones should be taken carefully. The
point is that any excited atomic state has a finite life time
(even in a cavity) and therefore the radiation is specified by a
certain line width that should be taken into account.

At far distances, we have
\begin{eqnarray}
j_{\ell} \sim \frac{1}{x} \sin (x- \ell \pi /2), \quad x=kr \gg
\ell , \nonumber
\end{eqnarray}
so that
\begin{eqnarray}
\vec{S}(kr) = \vec{L}(kr) \sim \frac{\hbar}{2V} \frac{\sin^2
(kr)}{(kr)^2} \vec{J} , \quad kr \gg 2. \label{12}
\end{eqnarray}
Thus, the spin and OAM densities contribute equally into the total
AM of a monochromatic $E1$ photon in the wave zone. Because of the
same operator structure, it is impossible to distinguish between
the spin and OAM parts by any measurement in the wave zone. This
result reflects well known fact that the total AM of $E1$ photon
cannot be divided into the spin and OAM parts \cite{13}.

Consider now emission of a photon by the same atomic transition as
above but take into consideration the natural line breadth. The
time-dependent wave function of the atom-field system can be
represented as follows
\begin{eqnarray}
| \psi(t) \rangle = C| \psi^{(0)} \rangle + \int B(k,t)| \psi (k)
\rangle dk, \label{13}
\end{eqnarray}
where the first term corresponds to the excited atomic state and
the vacuum state of the field. In turn, the second term gives the
ground atomic state and a single $E1$ photon. Employing the Markov
approximation then gives
\begin{eqnarray}
C(t) & = & e^{-i \omega_0 t- \Gamma t}, \nonumber \\ B(k,t) & = &
- \frac{k^{3/2}}{\omega_k - \omega_0 +i \Gamma} \left( 1-e^{i(
\omega_k- \omega_0)t - \Gamma t} \right), \nonumber
\end{eqnarray}
where $\omega_0$ is the atomic transition frequency and $\Gamma$
is the radiative decay width. Through the averaging of the
operators (2) and (3) in the spherical wave approximation (6) over
the state (13), we get
\begin{eqnarray}
\langle S_z (t) \rangle = \langle L_z(t) \rangle = \frac{\hbar}{2}
(1-e^{-2 \Gamma t}). \label{14}
\end{eqnarray}
Since the Markov approximation implies the long-time scale $t \geq
1/ \Gamma$ \cite{20}, the last result (14) corresponds to the
distances $r \geq c/ \Gamma \gg c/ \omega_0$ that again agree with
the wave zone. This means that the spin and OAM parts contribute
equally into the total AM in the wave zone independent of whether
or not we take the natural line breadth into account. Effects in
the near zone deserve special consideration.

Consider now quantum fluctuations of AM of $E1$ photon. Assume
that the atomic transition emits a single photon in the Fock state
$|1_m \rangle$, where $m=0, \pm 1$. It is then seen that
\begin{eqnarray}
\langle ( \Delta J_{x,y})^2 \rangle = \left\{ \begin{array}{ll} 1,
& \mbox{if $m=0$} \\ \frac{1}{2} , & \mbox{otherwise} \end{array}
\right. \label{15}
\end{eqnarray}
For all $m$, we have $\langle ( \Delta J_z)^2 \rangle =0$. Thus,
the state of $E1$ photon with projection $m=0$ of the total AM
undergoes more strong quantum fluctuations than the states with
$m= \pm 1$. This means that the Fock state $|1_0 \rangle$ of $E1$
photon should be less stable than $|1_{\pm} \rangle$.

At first sight, it may seem strange that the spin density operator
of $E1$ photon in (8) has all three components. The polarization
is usually associated with spin states and only two polarizations
are allowed for a photon \cite{11,13}. In fact, there is no
contradiction. The point is that the components of AM operator (9)
are determined in a reference frame connected with the atomic
dipole moment and that the Poynting vector not necessarily
coincides with the radial direction at any point. At $r
\rightarrow 0$, Poynting vector vanishes, so that $E1$ photon is
created with three polarizations (two circular and one linear
along the $z$-axis) \cite{21}. At any $r>0$, a local
transformation of the reference frame, turning the $z$-axis in the
direction of Poynting vector, can be determined \cite{9}. In this
local frame, there are only two polarizations. For the definition
of polarization operators in the case of multipole radiation, see
Ref. \cite{22}.

For the purposes of quantum information processing, the emission
of photon twins by a dipole forbidden atomic transition is of high
interest because the photons can be entangled in this case (e.g.,
see \cite{10}). In the usual treatment, the polarization
entanglement is considered. Here we examine the entanglement with
respect to AM.

Assume that the excited state has the AM $j_e=2$, while for the
ground state $j_g=0$. The two photons emitted by the transition
take away the angular momentum $j=2$. Thus, according to the
classification of two-photon states \cite{13}, the radiation field
can be observed in three states. Two of them are even and one is
odd. Since the excited atomic state is fivefold degenerated with
respect to the projection of AM $m$, assume that the photon twins
are created by the transition
\begin{eqnarray}
|j_e=2,m_e=0 \rangle \rightarrow |j_g=0,m_g=0 \rangle . \label{16}
\end{eqnarray}
Then, in view of the conservation of the total projection, the
even states are
\begin{eqnarray}
|\psi_1 \rangle = |1_0;1_0 \rangle , \nonumber \\  |\psi_2 \rangle
= \frac{1}{\sqrt{2}} (|1_+;1_- \rangle +|1_-;1_+ \rangle ),
\nonumber
\end{eqnarray}
while the odd state is
\begin{eqnarray}
|\psi_3 \rangle = \frac{1}{\sqrt{2}} (|1_+;1_- \rangle -|1_-;1_+
\rangle ). \nonumber
\end{eqnarray}
Here $|1_m;1_{m'} \rangle =|1_m \rangle \otimes |1_{m'} \rangle$
and the two photons move in opposite directions. The state
$|\psi_3 \rangle$ is the eigenstate of the Hamiltonian of
atom-photon interaction
\begin{eqnarray}
H= \sum_m \omega a^+_ma_m + \omega_0 R_{ee} \nonumber \\ + \gamma
\sum_{m,m'} (R_{eg}a_ma_{m'}+a^+_ma^+_{m'}R_{ge}) \label{17}
\end{eqnarray}
and therefore cannot be achieved in the process of radiation.
Summation in the last term in (17) is performed under the
condition $m+m'=0$ and $R_{ij}=|i \rangle \langle j|$ is the
atomic operator. Hence, the eigenstate of the radiation field can
be chosen  in the following form
\begin{eqnarray}
| \psi \rangle = \psi_1 |\psi_1 \rangle + \psi_2| \psi_2 \rangle,
\label{18}
\end{eqnarray}
By construction, (18) represents a nonseparable two-qutrit state
and may manifest entanglement \cite{23}. Since the measure of
entanglement in this case is $\mu ( \psi )=|\psi_1 \psi_2^2|$
\cite{24}, the state (18) is entangled if $\psi_1,\psi_2 \neq 0$.
To find the state, corresponding to maximum entanglement, we
should use the variational principle of Ref. \cite{24}. For
qutrits with the $SU(3)$ dynamic symmetry of the Hilbert space,
the measurements that can be performed over photons are provided
by the Hermitian generators of the $SU(3)$ subalgebra in the
Weyl-Heisenberg algebra (7). These generators are represented as
follows
\begin{eqnarray}
[M]= \left( \begin{array}{c} a^+_ma_m-a^+_{m-1}a_{m-1} \\
\frac{1}{2} (a^+_ma_{m-1}+H.c.) \\ \frac{1}{2i}
(a^+_ma_{m-1}-H.c.) \end{array} \right) \label{19}
\end{eqnarray}
Here we assume cyclic permutation of the subscript $m$, so that
$m-1=+1$ if $m=-1$. Among the three operators in the first row in
(19) only two are independent.

The variational principle for maximum entanglement \cite{24} can
be expressed by the condition \cite{25,26}
\begin{eqnarray}
\forall M \quad \langle \psi|M|\psi  \rangle =0, \nonumber
\end{eqnarray}
which is valid if
\begin{eqnarray}
|\psi_2|= \sqrt{2} |\psi_1|= \sqrt{2/3}. \label{20}
\end{eqnarray}
Thus, the transition (16) gives rise to photon twins entangled
with respect to the projection of AM. The two-qutrit state (18)
can be maximum entangled under the condition (20). Let us stress
that usually the two-qubit entangled state with respect to
polarization is considered \cite{10}.

Summarizing, we have discussed the spin and OAM parts of the total
AM of $E1$ photons emitted by atoms. We showed that the use of the
representation of spherical waves of photons leads to a violation
of commutation relations (4) for the spin and OAM. The physical
result coming from Eqs. (12) and (14) is that the spin and OAM
cannot be distinguished by means of any measurement in the wave
zone. The use of an idealized single-mode model shows the
difference in the distance dependence between spin and OAM in the
near zone. We showed that the two electric dipole photons emitted
by a dipole forbidden transition can manifest entanglement and
even maximum entanglement with respect to the projection of AM.

The obtained results can be generalized on the case of other
multipole photons. In particular, the case of magnetic dipole
photons is of some considerable interest because the photon states
are specified in this case by a unique OAM quantum number
\cite{13}.

\end{document}